\documentclass[pdflatex,sn-nature]{sn-jnl}

\usepackage{graphicx}
\graphicspath{{figure/}{SI_figures/}}
\usepackage{multirow}
\usepackage{amsmath,amssymb,amsfonts}
\usepackage{amsthm}
\usepackage[title]{appendix}
\usepackage{xcolor}
\usepackage{textcomp}
\usepackage{manyfoot}
\usepackage{booktabs}
\usepackage{algorithm}
\usepackage{algorithmicx}
\usepackage{algpseudocode}
\usepackage{listings}
\usepackage{needspace}
\usepackage{float}
\usepackage{placeins}
\usepackage{bookmark}

\setcitestyle{super,open={},close={},sort&compress}

\theoremstyle{thmstyleone}

\theoremstyle{thmstyletwo}

\theoremstyle{thmstylethree}

\begin{document}

\title{Effective Graph Neural Representation of Chain Architectures of Block Copolymers Tested by Predicting RPA Instability Limits}

\author[]{\fnm{Xiyue} \sur{Zhang}}
\author[]{\fnm{Yutong} \sur{Wu}}
\author*[]{\fnm{Yi-Xin} \sur{Liu}\textsuperscript{*}}\email{lyx@fudan.edu.cn}

\affil[]{\orgdiv{State Key Laboratory of Molecular Engineering of Polymers, Department of Macromolecular Science}, \orgname{Fudan University}, \orgaddress{\city{Shanghai}, \postcode{200433}, \country{China}}}

\abstract{Advances in polymer synthesis and computational modeling are expanding the block-copolymer architecture space accessible for phase-behavior design. Machine learning can accelerate this exploration, yet a common representation capable of learning across diverse chain topologies remains lacking. Here, we present a graph neural network (GNN) representation of acyclic AB multiblock architectures, tested by random-phase approximation (RPA) instability prediction. Each chain is represented as a graph, enabling a common model to learn across diverse topologies. This representation, which organizes RPA-related information, retains predictive capability when extrapolated to architectures with more blocks and demonstrates its potential for reuse across tasks through transfer to instability-wavenumber prediction. By effectively linking block-copolymer chain architecture to field-theoretic response, this graph representation not only reconstructs stability-limit curves but also provides a practical route for future self-consistent field theory exploration of ordered phase space across complex topologies.}
\maketitle

\section{Introduction}\label{introduction}

Block copolymers (BCPs) form periodic nanostructures, and their phase behavior depends on segregation strength, composition, and chain architecture \cite{matsenUnifyingWeakStrongSegregation1996,matsenEffectArchitecturePhase2012,bates50thAnniversaryPerspective2017}. Advances in polymer synthesis and field-theoretic simulation now allow phase behavior to be investigated across increasingly diverse chain topologies \cite{aroraBroadlyAccessibleSelfConsistent2016,xuArchitecturalDesign2024}. Phase behavior has nevertheless generally been mapped for selected architectures along limited parameter-space slices \cite{jiangEffectsChainTopology2018,xieSingleGyroid2022}. Yet each additional block or junction introduces new connectivity and length parameters, causing the combined space of chain architectures and thermodynamic parameters to expand rapidly. Within predefined chain architectures, machine-learning methods have accelerated sampling of thermodynamic parameter spaces and the discovery of self-assembled phases \cite{zhaoAutonomousConstructionPhase2021,dongAIdrivenAutomatedConstruction2025a}. Extending this capability across architectures requires a representation that can carry information learned from calculated chains to the responses of uncalculated ones.

Graph-based methods have begun to automate field-theoretic calculations across complex chain architectures. Our previous graph-enhanced SCFT framework represented chains as junction--block graphs, enabling automated exploration that identified a previously unreported stable phase \cite{zhangAutomatedChainArchitecture2024}. Yong and Kim proposed a SMILES-inspired line notation for propagator dependencies \cite{yongDynamicProgrammingChain2025}, while equivalent-branch aggregation and graph-isomorphism-based subtree analysis exploit structural equivalence to reduce repeated calculations \cite{liEfficientComputationSingleChain2025,liuAcceleratingFieldbasedSimulations2026}. However, these methods primarily organize propagator calculations within individual architectures or reuse structurally equivalent branches; responses for nonequivalent architectures must still be solved separately, limiting efficient exploration of broader architecture spaces.

In parallel, polymer informatics has developed learning methods that convert polymer descriptions into predictive representations. Fixed descriptors and fingerprints provide an established basis for polymer-property prediction \cite{stuartSizingFeatureDescriptors2023,kimPolymerGenomeDataPowered2018,kuennethCopolymerInformaticsMultitask2021}. Sequence learning offers a natural extension of serialized chain descriptions such as Yong and Kim's SMILES-inspired notation \cite{yongDynamicProgrammingChain2025}. In related work, BigSMILES provides a line notation for stochastic polymers, TransPolymer learns property representations from polymer sequences, and G2RINS extends string-based representation to complex hierarchical architectures \cite{linBigSMILESStructurallyBasedLine2019,xuTransPolymerTransformerbasedLanguage2023,zaldivarG2RINSGenerativeStringandGraph2026}. The junction--block graph used in our previous SCFT framework provides a corresponding basis for graph learning \cite{zhangAutomatedChainArchitecture2024}. Evidence for the potential of this approach comes from graph neural networks (GNNs) developed by Jiang and co-workers. These networks were trained to predict the squared radius of gyration and its distribution from molecular dynamics simulations, and the learned representations transfer across polymer topologies \cite{jiangPropertyguidedGenerationComplex2024,jiangPhysicsGuidedNeuralNetworks2025}. Nevertheless, existing polymer representations have primarily been developed for property prediction, structure generation or conformational observables, highlighting the need for chain representations designed to predict architecture-dependent field-theoretic responses.

Here, we aim to learn an architecture representation that reuses response information across chain architectures while retaining physically meaningful structural information. To establish a physics-grounded learning benchmark over a broad range of chain architectures, we use the random-phase approximation (RPA). Its predictions of disordered-state instability depend on intramolecular correlations organized by blocks, shared junctions and intervening contour paths. The critical segregation parameter \((\chi N)^*\) and the corresponding dimensionless wavenumber \(k^*\) at the spinodal therefore provide architecture-sensitive prediction targets \cite{leiblerTheoryMicrophaseSeparation1980}. For prediction of the RPA instability limit \((\chi N)^*\), we compare three representations of chain architecture: an MLP acting on fixed descriptors, a transformer encoding canonical chain sequences and a topology-aware GNN mapping junction--block graphs to fixed-dimensional vectors (Fig. 1). The GNN provides the best overall balance between interpolation within the training architecture range and extrapolation to larger chain graphs. To investigate the basis of this extrapolation, layerwise diagnostics track how RPA-related structural information develops through message passing. We next transfer the \((\chi N)^*\)-pretrained encoder to \(k^*\) prediction to test whether the learned architecture representation can be reused across related targets and how that reuse depends on target-label availability, following the broader use of transfer learning in materials-property prediction \cite{guptaStructureawareGraphNeural2024}. Evaluating the model across composition produces architecture-dependent spinodal curves that locate the RPA instability limit.

\begin{figure}[t]
\centering
\includegraphics[width=\linewidth]{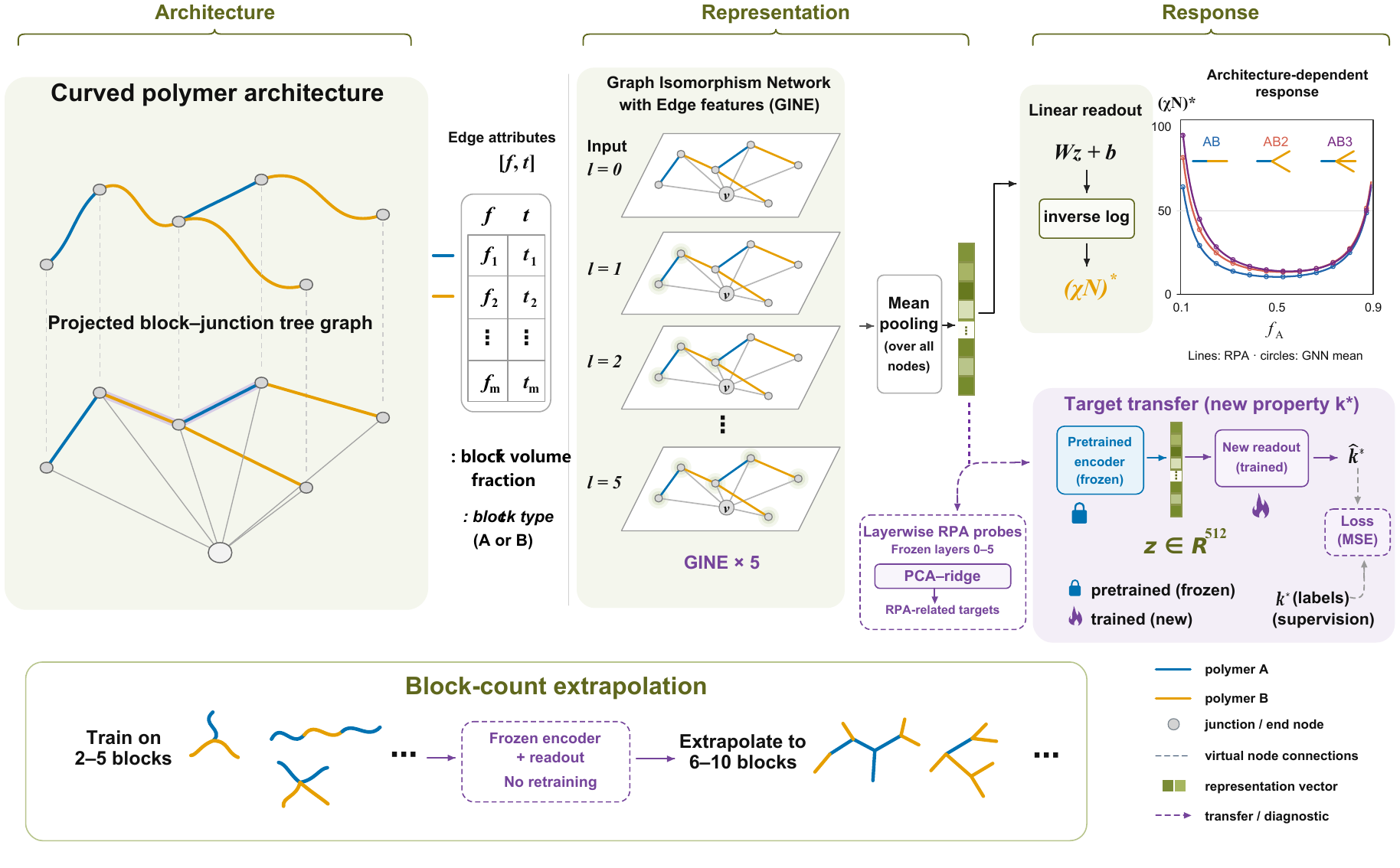}
\caption{\textbf{Graph neural representation of block-copolymer architectures tested by RPA instability prediction.} Left: chain architectures are encoded as block--junction graphs. Blue and orange edges represent A and B blocks, with volume fraction \(f\) and species label \(t\) as edge attributes. Grey circles denote junctions and chain ends, and an auxiliary virtual node enables global communication. Centre: a graph neural network (GNN) encoder uses five layers of a graph isomorphism network with edge features (GINE), followed by global mean pooling, to produce a 512-dimensional architecture vector. Upper right: a linear prediction head and inverse logarithmic transformation predict the critical segregation parameter \((\chi N)^*\). Solid curves show random-phase approximation (RPA) calculations, and circles show mean predictions from five GNN models trained with different random seeds, plotted against A volume fraction \(f_A\). \(AB\) denotes a linear diblock, while \(AB_2\) and \(AB_3\) denote stars with one A arm and two or three B arms, respectively. Below the encoder: principal component analysis (PCA) followed by ridge regression probes RPA-related information in frozen representations at layers 0--5. Lower right: the pretrained encoder is frozen and a new prediction head is trained to predict \(k^*\), the dimensionless wavenumber at the instability, using mean squared error (MSE) loss. Lock and flame symbols indicate frozen and trainable components, respectively. Purple dashed arrows mark diagnostic or transfer paths. Bottom: the frozen encoder and prediction head, trained on two- to five-block architectures, are evaluated on six- to ten-block architectures.}\label{fig:fig1}
\end{figure}

\FloatBarrier

\section{Results}\label{results}

\subsection{Architecture and response distributions}\label{architecture-and-response-distributions}

Before comparing architecture representations, we first established the structural and response domains covered by the data. The random topology generator produced a development pool of 100,000 AB architectures containing two to five blocks, with 25,000 architectures at each block count. Each architecture was labelled by its RPA instability threshold \((\chi N)^*\). A block-count-balanced test set of 2,000 architectures, with 500 examples at each block count, was used for final in-domain evaluation of \((\chi N)^*\) and \(k^*\).

Across these two- to five-block data, the sampled architectures span linear chains, single-junction branches and multi-junction trees (Fig. 2a). Increasing the block count introduces more connection and branching patterns, producing variation in graph size, topology and composition. The \(\log_{10}(\chi N)^*\) distributions shift with block count, but the response ranges covered by the development pool and test set overlap within each block-count group (Fig. 2b). The test set therefore evaluates interpolation to unseen architectures within the represented block-count and response ranges.

\begin{figure}[t]
\centering
\includegraphics[width=110mm]{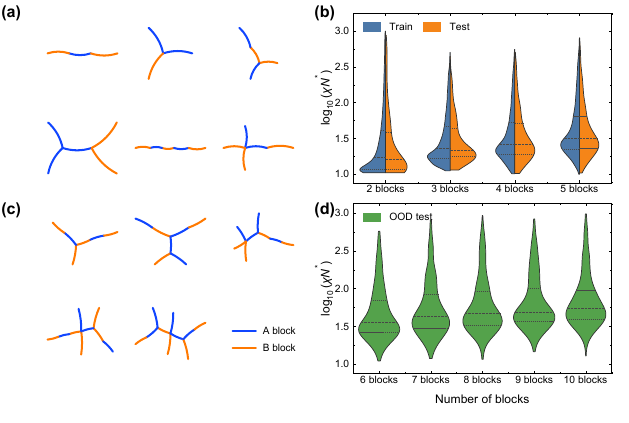}
\caption{\textbf{Polymer architecture and response distributions across block counts.} \textbf{a,} Representative randomly generated AB block-copolymer architectures from the two- to five-block dataset, spanning linear chains, single-junction branches, and multi-junction trees; blue and orange curves denote A and B blocks, respectively. \textbf{b,} Block-count-resolved distributions of \(\log_{10}(\chi N)^*\) in the 100,000-sample development pool and the 2,000-sample test set. Blue and orange halves denote development and test data, respectively. \textbf{c,} Representative OOD architectures containing six to ten blocks; graph size and accessible branching patterns increase with block count. \textbf{d,} Block-count-resolved distributions of \(\log_{10}(\chi N)^*\) in the independently generated OOD set, with 500 architectures per block count. Violin width represents probability density; internal horizontal lines mark the 25th, 50th (median), and 75th percentiles.}\label{fig:fig2}
\end{figure}

We constructed a separate out-of-distribution (OOD) dataset to test whether models trained on two- to five-block architectures can extrapolate to larger and more structurally complex graphs. This dataset contains 2,500 independently generated acyclic AB architectures, with 500 examples at each block count from six through ten. It follows the same sampling and RPA-labelling logic as the development data. The resulting graphs are larger and include more highly branched trees, while the number of accessible connection patterns increases with block count (Fig. 2c).

The response distributions change smoothly at the boundary between the training and OOD block counts (Figs. 2b and 2d). The upward shift in median \(\log_{10}(\chi N)^*\) observed from two to five blocks continues across the OOD set, rising from 1.55 at six blocks to 1.74 at ten blocks. The violin distributions remain broad and substantially overlapping, and the same response-acceptance criterion bounds both datasets to a common interval. The OOD task therefore varies architecture size and complexity while holding the response range fixed, isolating generalization across architectures.

\subsection{Architecture representations and in-domain prediction}\label{architecture-representations-and-in-domain-prediction}

To identify an effective representation of block copolymer chain architecture, we compared three ways of encoding the same chain for prediction. The descriptor MLP compresses each architecture into a set of global structural summaries. The canonical Transformer instead reads a sequence that preserves the attributes of individual blocks and their shared junctions, while the proposed GNN operates directly on the field-theoretic junction--block graph of Zhang \emph{et al}. \cite{zhangAutomatedChainArchitecture2024}.

All three models were tuned under a common procedure (Methods). Model-specific screening, hyperparameter selection and ablation results are reported in Supplementary Discussion Sections 6 to 8, Supplementary Tables S6 to S12 and Supplementary Figs. S12 to S21. The tuned models were then evaluated across five runs on the same 2,000-architecture test set.

\begin{figure}[b]
\centering
\includegraphics[width=\linewidth]{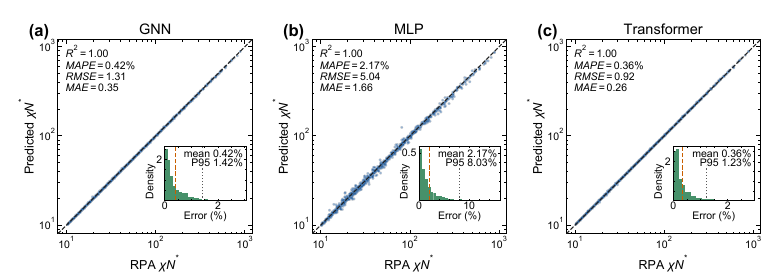}
\caption{\textbf{In-domain prediction by the three architecture representations.} Predictions from the common seed 202604 models for \textbf{a,} a graph neural network (GNN), \textbf{b,} a descriptor multilayer perceptron (MLP) and \textbf{c,} a canonical transformer on the same independent 2,000-architecture test set. Blue points compare predicted and RPA-derived \((\chi N)^*\) values, both axes are logarithmic and the dashed diagonal denotes exact agreement. Insets show the probability-density distributions of absolute percentage errors; orange dashed and grey dotted lines mark the mean and 95th percentile, respectively. Main-panel annotations report MAPE, RMSE, MAE and \(R^2\). Predictions from all five seeds are shown in Supplementary Fig. S1.}\label{fig:fig3}
\end{figure}

All three model families predicted the in-domain RPA response, while the graph and sequence models reached the same high-accuracy regime. Across five runs, the MAPE (mean ± standard deviation) was 2.21 ± 0.10\% for the descriptor MLP, 0.42 ± 0.09\% for the GNN and 0.37 ± 0.03\% for the canonical transformer. The GNN and transformer were substantially more accurate than the descriptor MLP and gave closely comparable in-domain performance. Complete metrics for all five runs are reported in Supplementary Table S1.

The matched-seed parity plots follow the same pattern (Figs. 3a to 3c). Predictions from all three models lie close to the parity line across the in-domain range, with greater scatter for the descriptor MLP. More parity plots are provided in Supplementary Fig. S1, and the corresponding signed relative-error distributions are shown in Supplementary Fig. S2. Across all three model families, training and validation losses also declined together, with no sustained separation late in training (Supplementary Figs. S3 to S5).

This performance ordering is consistent with the information retained by the three representations. The RPA instability limit depends on intramolecular correlations determined by block identity and fraction, shared junctions and the contour paths connecting different blocks. The graph and canonical sequence representations retain these relations at the level of individual blocks, whereas the descriptor representation summarizes them through a fixed set of aggregate quantities.

\subsection{Block-count extrapolation}\label{block-count-extrapolation}

The in-domain comparison does not establish whether the three architecture representations remain predictive beyond the block-count range used for training. We therefore applied the trained GNN, transformer and descriptor MLP, with all parameters frozen, to independently generated six- to ten-block architectures. Figures 4a and 4b resolve this comparison by block count.

\begin{figure}[t]
\centering
\includegraphics[width=\linewidth]{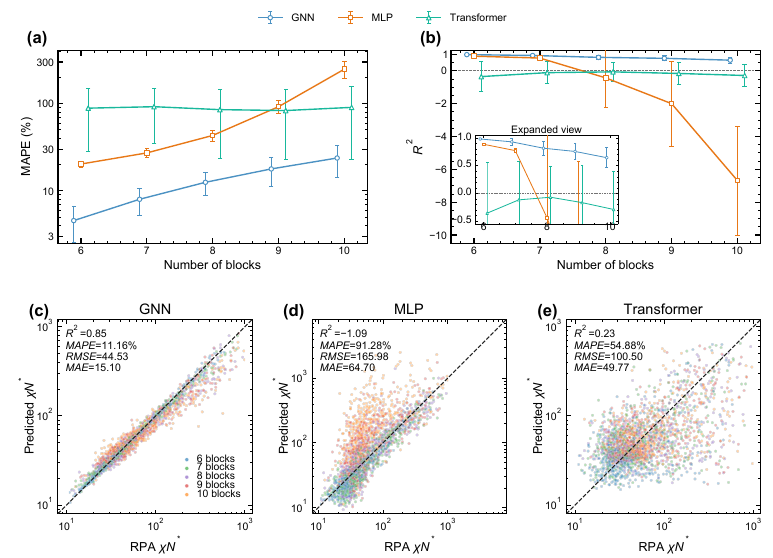}
\caption{\textbf{Block-count extrapolation of the three architecture representations.} Five-seed mean \textbf{a,} MAPE and \textbf{b,} \(R^2\) for the graph neural network (GNN), descriptor multilayer perceptron (MLP) and canonical transformer on independently generated six- to ten-block architectures, with 500 samples at each block count. Panel a uses a logarithmic vertical axis. Error bars show the sample standard deviation across the five frozen models. The horizontal dashed lines in b mark \(R^2=0\); the inset expands the range \(-0.55\leq R^2\leq1.05\). \textbf{c--e,} Predictions from the common seed 202604 model for \textbf{c,} GNN, \textbf{d,} descriptor MLP and \textbf{e,} canonical transformer on the same 2,500 architectures. Seed 202604 had the closest joint match to the median performance rank across the three model families. Colours identify block count, both axes are logarithmic and the dashed diagonal denotes exact agreement. Panel annotations report \(R^2\), MAPE, RMSE and MAE over the complete extrapolation set; all-seed panels are provided in Supplementary Fig. S6.}\label{fig:fig4}
\end{figure}

Across five runs on the complete OOD set, the GNN achieved MAPE 13.33 ± 4.74\% and \(R^2\) = 0.80 ± 0.11. The corresponding values were 86.26 ± 15.26\% and -2.03 ± 1.40 for the descriptor MLP, and 88.02 ± 61.46\% and -0.17 ± 0.67 for the transformer. Thus, despite their similarly high in-domain accuracy, the GNN and transformer behaved very differently on unseen block counts. Complete MAPE, RMSE, MAE and \(R^2\) values are reported in Supplementary Tables S2 and S3.

The block-resolved results show how this separation developed. For the GNN, mean MAPE increased from 4.55 ± 1.99\% at six blocks to 23.77 ± 9.45\% at ten blocks, while mean \(R^2\) decreased from 0.98 ± 0.02 to 0.64 ± 0.18. The descriptor MLP began at 20.28 ± 1.64\% MAPE for six blocks and reached 248.53 ± 54.54\% for ten blocks, with mean \(R^2\) becoming negative from eight blocks onward. The transformer MAPE remained between 83.30\% and 92.21\% across all five unseen block counts, and none of these groups had a positive mean \(R^2\). Over the familiar response scale established in Fig. 2, the GNN retained useful predictive ability throughout the tested block counts, whereas both baselines lost most of their in-domain accuracy as architecture size increased.

The combination of high in-domain accuracy and a clear extrapolation advantage indicates that the GNN provides an effective representation of block-copolymer chain architecture for predicting RPA instability limits within the block-count range studied here. We next asked why this representation works: which physically meaningful quantities it encodes, and whether that encoded content can support prediction beyond the observable used for training.

\subsection{GNN feature impact in relation to RPA correlations}\label{gnn-feature-impact-in-relation-to-rpa-correlations}

To understand how the GNN uses chain architecture, we examined the five-block topology \(A_1B_1(B_2,B_3)_TA_2\) (Fig. 5a). This architecture contains within-block, shared-junction and separated-path correlations, providing a simple setting in which to relate the learned representation to RPA. Keeping the connectivity fixed, we varied the overall A fraction \(f_A\) and the relative fractions of the B blocks \(\rho\) to obtain 45 parameterizations spanning symmetric and increasingly asymmetric block allocations. The exact values and block-fraction definitions are given in Methods.

At the level of individual blocks, we first asked whether RPA influence is associated with the GNN's prediction sensitivity. We compared the sensitivity-weighted RPA contribution assigned to each edge, \(I_e^{\rm RPA}\), with the prediction change \(F_e\) when messages along that edge were suppressed (Methods). In the representative symmetric parameterization, the three RPA components partly reinforce and partly cancel one another (Fig. 5b), and their combined influence is compared with model sensitivity in Fig. 5c. The pooled correlation across 225 edge-level comparisons was \(\rho_s=0.644\), and the within-parameterization association was positive in 43 of the 45 parameterizations (Supplementary Fig. S7a). The GNN's edge sensitivity is therefore positively associated with the RPA edge influence across the composition and asymmetry range.

\begin{figure}[t]
\centering
\includegraphics[width=\linewidth]{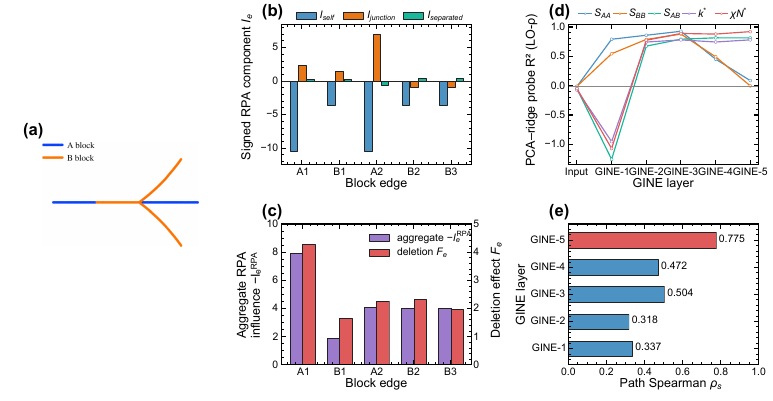}
\caption{\textbf{GNN feature impact in relation to RPA correlations.} \textbf{a,} Diagnostic five-block topology \(A_1B_1(B_2,B_3)_TA_2\); blue and orange segments denote A and B blocks, respectively, and the subscript \(T\) marks the branch junction. \textbf{b,} Edge-resolved RPA self, shared-junction and separated-path components for the symmetric instance at \(f_A=0.5\) and \(\rho=1\). \textbf{c,} Aggregated RPA edge influence \(-I_e^{\rm RPA}\) and mean physical-message deletion effect \(F_e\) for the same instance; the two quantities retain separate vertical scales. \textbf{d,} Grouped-by-\(\rho\) five-fold PCA--ridge-probe \(R^2\) for \(S_{AA}(k^*)\), \(S_{BB}(k^*)\), \(S_{AB}(k^*)\), \(k^*\) and \((\chi N)^*\) from the pooled input representation and layers 1--5 of the graph isomorphism network with edge features (GINE). \textbf{e,} Spearman association between layerwise embedding distance and RPA-vector distance across 16,065 block-placement counterfactual pairs after averaging embeddings over all legal growth histories; the input representation is omitted because its within-group distance is constant and the correlation is undefined. Panels b and c show the representative symmetric instance, whereas panels d and e summarize the complete analysis grid.}\label{fig:fig5}
\end{figure}

Beyond output sensitivity, we examined which RPA quantities could be read from successive encoder layers. Using probes that combined PCA with ridge regression and held out complete B-block asymmetry groups for evaluation (Methods), we predicted \(S_{AA}(k^*)\), \(S_{BB}(k^*)\), \(S_{AB}(k^*)\), \(k^*\) and \((\chi N)^*\) from each layer. All five probe scores were negative at the pooled input and became positive by layer 2 (Fig. 5d). The correlation functions were best predicted from layers 3--4, reaching \(R^2=0.927\), 0.879 and 0.815 for the AA, BB and AB channels, respectively. The derived quantities \(k^*\) and \((\chi N)^*\) were best predicted from the final layer, with \(R^2=0.780\) and 0.921. This progression is consistent with the RPA relationship between intramolecular correlations and the instability threshold: correlation information is most accessible in the intermediate representations, whereas the final representation is most informative about the resulting spinodal quantities.

Finally, we asked whether the representation distinguishes where a fixed set of blocks is placed within the same topology. Within each parameter cell, we kept the topology, total composition and set of block attributes fixed and reassigned those attributes among the edges. This reassignment left the global block summaries unchanged but altered the shared-junction relations and contour separations that determine the RPA correlations. We tested the association between representation distance and the distance between RPA vectors containing separated-path correlations and \(k^*\). To account for node ordering, we averaged the representations over all legal growth histories. Details are provided in Methods, Supplementary Discussion Section 3, and Supplementary Fig. S7b. Across 16,065 matched distance pairs, the association was strongest at the final layer, with a Spearman correlation of 0.775 (Fig. 5e). Final-layer correlations were positive in all 45 parameter cells and at least 0.5 in 43 (Supplementary Fig. S7b). These results show that the final representation resolves block-placement differences and relates them to the corresponding changes in RPA correlation structure.

Together, the edge-influence, layerwise-probe and block-placement analyses provide complementary evidence that the GNN representation retains architecture information associated with the RPA response. Because \(k^*\) was recoverable from the final representation despite the encoder being trained only on \((\chi N)^*\), we next tested whether this cross-target information supported \(k^*\) prediction across the full two- to five-block architecture distribution.

\subsection{Cross-target transfer learning}\label{cross-target-transfer-learning}

We transferred the \((\chi N)^*\)-pretrained encoder to \(k^*\) prediction under two regimes of labelled training data for the downstream task. Reduced \(k^*\)-label sets tested whether architecture pretraining improved accuracy at a fixed labelled-data budget, whereas the full training pool was used to compare the degree of encoder adaptation required.

We first examined how pretraining affected prediction when labelled training data for the downstream task were limited. We compared the methods using 100, 250, 500, 1,000 or 2,000 labelled architectures, and every model was evaluated on the same independent test set (Supplementary Table S4a). For models trained with 100 labels, the MAPEs were 4.12 ± 1.12\% after end-to-end fine-tuning and 3.77 ± 0.56\% after two-stage transfer, compared with 10.58 ± 0.49\% for scratch training (Fig. 6a--e and Supplementary Fig. S8). Increasing the budget to 2,000 labels reduced the corresponding errors to 1.05 ± 0.16\%, 1.00 ± 0.07\% and 2.21 ± 0.26\%. Figure 6f shows the broader pattern across the five paired seeds. Both encoder adaptation protocols retained lower errors than scratch training at every subset size, showing that the benefit of pretraining persisted throughout the selected limited-data range.

\begin{figure}[!htbp]
\centering
\includegraphics[width=\linewidth]{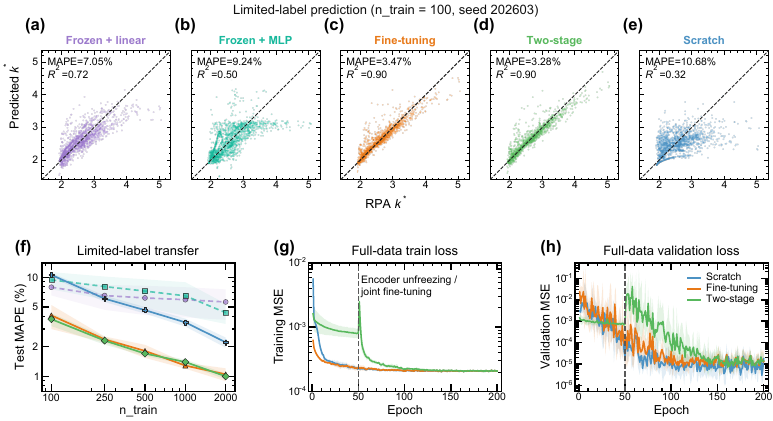}
\caption{\textbf{Transfer from \((\chi N)^*\) to \(k^*\) prediction.} \textbf{a--e,} Representative matched-seed parity plots for frozen linear decoding, frozen-embedding MLP decoding, end-to-end fine-tuning, two-stage transfer and scratch training using 100 \(k^*\)-labelled training architectures. Seed 202603 was selected by the smallest mean absolute standardized deviation from the five-seed mean across the five methods. All panels show predictions for the same independent, block-count-balanced test set of 2,000 architectures; the dashed diagonal denotes exact agreement. Panel annotations report test MAPE and \(R^2\). \textbf{f,} Test MAPE for paired, strictly nested and block-count-balanced subsets containing 100--2,000 \(k^*\)-labelled training architectures. Lines and shaded bands denote the mean and sample standard deviation across five seeds, respectively. \textbf{g,h,} Full-data optimization histories for scratch training, end-to-end fine-tuning and two-stage transfer using 80,000 training architectures. Panels \textbf{g} and \textbf{h} show training and validation MSE, respectively, in the \(\log_{10}(k^*)\) target space. Curves are geometric means across seeds 202601--202605; shaded bands represent one sample standard deviation after transforming the per-seed losses to \(\log_{10}(\mathrm{MSE})\). The vertical dashed line at epoch 50 marks encoder unfreezing in the two-stage protocol and the transition from decoder-only adaptation to joint encoder--decoder fine-tuning.}\label{fig:fig6}
\end{figure}

We next examined how much \(k^*\) information could be recovered without updating the pretrained encoder. With the full training pool, a linear decoder fitted to the frozen architecture vectors gave a MAPE of 3.66 ± 0.39\%, whereas a nonlinear decoder with two hidden layers reduced the error to 1.20 ± 0.40\% (Supplementary Table S4b). Thus, \(k^*\) was recoverable from the frozen representation, but a linear prediction head did not extract all of the predictive information available in the final architecture vector.

Allowing the encoder to adapt further reduced the full-data error. End-to-end fine-tuning and two-stage transfer reached MAPEs of 0.27 ± 0.07\% and 0.29 ± 0.04\%, respectively, whereas scratch training achieved the lowest error, 0.18 ± 0.04\% (Supplementary Table S4b). Thus, pretraining improved accuracy when \(k^*\) labels were limited, but it did not improve the best attainable accuracy when all 80,000 target-labelled architectures were available.

The transfer protocols nevertheless required less encoder adaptation. Immediate fine-tuning entered the low-training-loss regime within the first few epochs, whereas scratch training required more epochs to descend from its random initialization (Fig. 6g). In two-stage transfer, the encoder was updated only during the final 150 of 200 epochs, reducing encoder-update epochs by 25\% relative to immediate fine-tuning while retaining a MAPE below 0.30\%. Once the encoder was unfrozen, its training and validation losses approached those of immediate fine-tuning (Fig. 6g,h). Full-data parity plots and the corresponding signed relative-error distributions are provided in Supplementary Figs. S10 and S11.

\subsection{Stability-limit curves across composition}\label{stability-limit-curves-across-composition}

We next tested whether the learned field-theoretic graph representation can reconstruct composition-dependent RPA stability-limit curves across distinct chain architectures. Shi showed that block number, junction topology and block-length asymmetry reshape the scattering functions and spinodal transitions of linear and nonlinear AB-type block copolymers \cite{shiScatteringFunctionSpinodal2021}. The GNN provides a common learned mapping from graph topology and block fractions to the RPA stability limit across these architectures. A single model uses the same graph interface for each architecture and learns how its topology and block fractions shape the stability limit. Unless otherwise stated, predictions were evaluated at 100 uniformly spaced compositions over \(f_A\in[0.01,0.99]\). The RPA references were computed with Polyorder.jl. Supplementary Table S5 lists the topology definitions, block-fraction assignments and literature sources for the benchmark architectures.

We first used three benchmark architecture series to test whether the GNN captures how spinodal curves change with systematic variations in topology and block number. The \(AB\), \(AB_2\), and \(AB_3\) series lies entirely within the two- to five-block development domain (Fig. 7a). The alternating linear \(AB\), \(ABAB\), and \(ABABAB\) series extends the comparison to six-block topology (Fig. 7b), while the third series connects \(AB\) to the nonlinear \(B_2A_2\) architecture and its eight-block repeat \((B_2A_2)_2\) (Fig. 7c), following the architectures analyzed by Shi \cite{shiScatteringFunctionSpinodal2021}. These series therefore pair in-domain topology changes with block-count extrapolations beyond the training range. The agreement with independently computed RPA curves demonstrates that the model recovers the curve shapes, approximate minima and systematic topology-dependent shifts in \((\chi N)^*\) across all three series. The five-model mean curves for the two extrapolated architectures systematically underestimate the RPA values, with MAPEs of 7.45\% for six-block \(ABABAB\) and 12.45\% for eight-block \((B_2A_2)_2\), but retain their overall curve shapes, approximate minima and composition-dependent trends. The larger error for eight-block \((B_2A_2)_2\) than for six-block is consistent with the increase in extrapolation error observed at higher block counts.

\begin{figure}[!htbp]
\centering
\includegraphics[width=\linewidth]{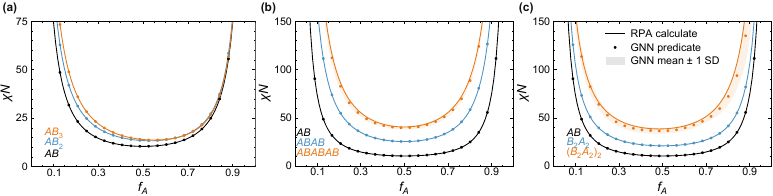}
\caption{\textbf{RPA stability-limit curves across benchmark architecture series.} \textbf{a,} \(AB\), \(AB_2\), and \(AB_3\). \textbf{b,} The linear \(AB\), \((AB)_2\), and \((AB)_3\) series. \textbf{c,} \(AB\), \(B_2A_2\), and eight-block \((B_2A_2)_2\). Solid curves show random-phase approximation (RPA) calculations, whereas circles and shaded bands show the mean and one sample standard deviation of graph neural network (GNN) predictions across five independently trained models, respectively.}\label{fig:fig7}
\end{figure}

\begin{figure}[H]
\centering
\includegraphics[width=\linewidth]{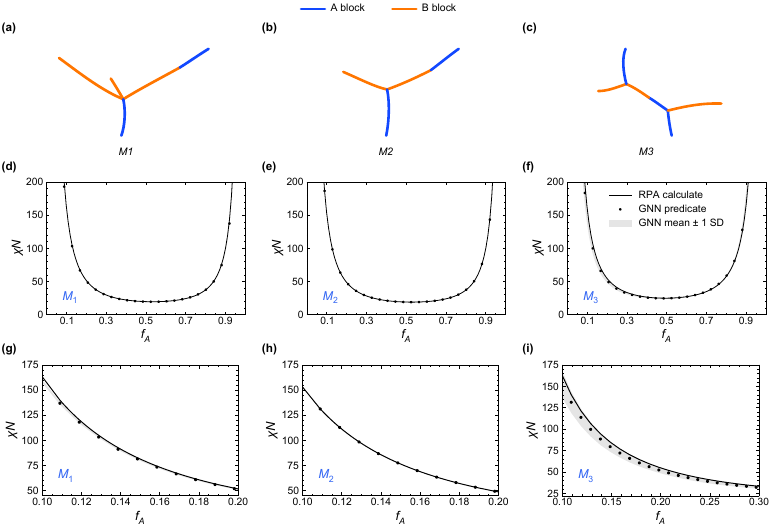}
\caption{\textbf{Architectures and RPA stability-limit curves for literature-derived cases.} \textbf{a--c,} Chain architectures \(M_1\), \(M_2\) and six-block \(M_3\), respectively; blue and orange segments denote A and B blocks. \textbf{d,g,} \(M_1\), with the local window \(f_A\in[0.10,0.20]\). \textbf{e,h,} \(M_2\), with the local window \(f_A\in[0.10,0.20]\). \textbf{f,i,} \(M_3\), with the local window \(f_A\in[0.10,0.30]\). Panels \textbf{d--f} show the global composition scans, whereas panels \textbf{g--i} enlarge the corresponding composition windows examined in the cited phase-diagram studies. The shared curve legend is shown in \textbf{f}. Solid curves show random-phase approximation (RPA) calculations, whereas circles and shaded bands show the mean and one sample standard deviation of graph neural network (GNN) predictions across five independently trained models, respectively.}\label{fig:fig8}
\end{figure}

After evaluating systematic topology and block-count series, we examined literature-derived architectures associated with specific phase-stability questions. Rapid reconstruction of the RPA stability limit can identify composition intervals in which the disordered state loses stability and guide subsequent ordered-phase calculations. We therefore selected literature architectures used to address specific phase-stability questions, including both cases within the training block-count range and a block-count extrapolation case. Our previous graph-enhanced SCFT work discovered \(M_1\) and \(M_2\), providing five- and four-block in-domain cases, respectively \cite{zhangAutomatedChainArchitecture2024} (Fig. 8a,b). We denote the six-block architecture \(A_1(B_1)_TB_2A_2(A_3)_TB_3\), designed by Xie et al. (2020) to study low-coordination spherical phases, as \(M_3\), which serves as a block-count extrapolation case (Fig. 8c) \cite{xieSynergisticEffectStretched2020}.

Across the global composition scans, the five-model mean curves reproduce the overall curvature and minima of \(M_1\), \(M_2\) and \(M_3\), with full-range MAPEs of 4.98\%, 3.50\% and 6.78\%, respectively (Fig. 8d--f). Because the rapid rise near the composition endpoints compresses the intervals examined in the cited phase-diagram studies, the local panels enlarge these regions to show the agreement more clearly (Fig. 8g--i). The \(M_1\) mean curve shows a small systematic underestimation while retaining the same decay, whereas \(M_3\) preserves the composition dependence with a larger but consistent offset. For the tested architectures, these results suggest that our trained GNN model is a fast and reliable tool for delineating the ODT phase boundaries.

\section{Discussion}\label{discussion}

An effective representation of block-copolymer architectures must jointly encode block species, contour lengths and junction-mediated chain connectivity, because these features determine the intramolecular correlations entering the RPA response. The descriptor MLP can encode this information only indirectly through preselected summaries of topology and composition. More training data improves prediction within that descriptor space (Supplementary Fig. S18), but it does not expose block connectivity and placement more explicitly to the model. Under matched training conditions, validation performance improved only when the representation retained block species and fractions, used the junction--block construction and included a global communication path (Supplementary Tables S7a--S7c and S8a--S8b).

The preceding ablations show that explicit connectivity improves in-domain validation performance, but they do not establish whether the encoded relations can be reused when architecture size exceeds the training range. The canonical transformer provides a useful contrast: its input sequence contains complete block and junction information by construction, and it achieves near-perfect in-domain accuracy comparable to that of the GNN. Exact serialization-invariance checks further confirm that equivalent sequence encodings produce identical predictions (Supplementary Discussion Section 8). Nevertheless, its predictions deteriorate on OOD architectures, indicating that the sequence model does not effectively reuse the encoded information beyond the training block-count range. This behavior is consistent with prior findings that transformer length extrapolation depends strongly on the positional-encoding design \cite{ruossRandomizedPositionalEncodings2023,kazemnejadImpactPositionalEncoding2023}. The GNN, by contrast, applies the same junction--block representation and shared message-passing operations as graph size increases: information propagates along physical connections, while the virtual node maintains a global communication path. The GNN's stronger OOD performance, together with the ablation results, suggests that this graph-based design enables connectivity information to be reused more effectively across architecture sizes.

Mechanistic analyses further clarify what the GNN retains in its architecture representation. The block-placement counterfactuals show that its encoder distinguishes architectures with the same composition and block-attribute multiset but different junction relations and contour separations. Together with the edge-influence and layerwise-probe results, these findings show that the GNN retains architecture information relevant to the intramolecular correlations entering RPA. The composition scans extend this evidence from discrete changes in topology to continuous changes in block fraction and show that the retained information remains predictive even for architectures beyond the training block-count range. Taken together, these results indicate that the GNN can use RPA-relevant architecture information across changes in topology, composition and graph size. Therefore, they suggest that a better training scheme would be to train on several block counts distributed across the intended application range and reserve the intervening counts for interpolation tests, thereby reducing reliance on one-sided extrapolation from only small architectures.

Prediction of \(k^*\) from the frozen encoder shows that the architecture vector can be reused beyond its original training target. The two targets depend on the same intramolecular correlations, but the encoder was pretrained using only the \((\chi N)^*\) target. The lower error obtained with a nonlinear decoder further indicates that the frozen vector contains predictive information that a linear prediction head does not fully recover. With 100--2,000 \(k^*\) labels, encoder fine-tuning and two-stage transfer outperformed scratch training for every paired seed, demonstrating an accuracy advantage under limited target data. With the full training pool, scratch training achieved the lowest error. Frozen decoding restricted optimization to the target decoder, whereas two-stage transfer used 25\% fewer encoder updates than immediate fine-tuning, but both incurred higher prediction error than scratch training. The value of representation reuse therefore depends on the target-data regime: it improves predictive accuracy when \(k^*\) labels are limited and reduces parameter or encoder updates when target data are abundant. This regime dependence is directly relevant to future transfer from RPA to SCFT, for which labelled data are computationally expensive to generate. A pretrained architecture encoder could be adapted when only a limited number of SCFT labels are available, using frozen or staged protocols when encoder updates are constrained, before shifting to full SCFT-specific training if a sufficiently large labelled dataset becomes available.

The extrapolation, interpretability and target-transfer results support the GNN as an effective representation of polymer architecture for RPA prediction. RPA serves here as a controlled physical benchmark for the representation. The prospective computational payoff is its reuse for more expensive SCFT targets, which remains to be tested. The framework is expected to work on SCFT because RPA and SCFT share the same single-chain statistics. It is worth noting that extending the current graph input to SCFT requires adding \(\chi N\) to the block-edge features so that the representation is conditioned on segregation strength. To test this transfer, a new decoder would be attached to the RPA-pretrained encoder, and the model would be fine-tuned on phase-specific SCFT free energies.

A further extension could combine the architecture representation with a conditional variational autoencoder (VAE) and a graph decoder to generate candidate topologies for a target morphology. Free-energy predictions could screen these candidates, followed by SCFT verification of their phase stability. This would extend the framework from predicting architecture-dependent responses to inverse design of polymers that stabilize a desired phase.

\section{Methods}\label{methods}

\subsection{Dataset construction, partitioning and evaluation protocol}\label{dataset-construction-partitioning-and-evaluation-protocol}

We generate AB copolymer architectures by sampling block connectivity, species and volume fractions. Each architecture is represented as an acyclic undirected graph, with blocks as edges and junctions or free ends as vertices. The resulting dataset includes linear, star and multiply branched chains containing both A and B blocks. The same generation and RPA-labelling workflow extends from two to five blocks for model development to six to ten blocks for OOD evaluation. Exact sampling and numerical settings are provided with the accompanying code.

Architectures are represented and processed using Polymer.jl v0.9.1, and the RPA labels are calculated using Polyorder.jl v0.24.3. Compositions are sampled away from the pure-component limits and distributed among the constituent blocks. RPA calculations supply the instability threshold \((\chi N)^*\) and corresponding dimensionless wavenumber \(k^*\). The accepted architectures and labels form the common data source for the GNN, transformer and MLP.

The block-count-balanced development pool contains 100,000 two- to five-block architectures and is partitioned into 80,000 training samples and 20,000 internal-validation samples. Only the training samples contribute to gradient updates. The internal-validation samples are used for model and hyperparameter selection. For the primary \((\chi N)^*\) task, training-size analyses use strictly nested subsets spanning 8,000 to 80,000 samples from the fixed training partition. An independently generated validation set provides a common basis for comparison across training sizes. Its construction and the protocol-specific checkpoint procedures are described in the Supplementary Information.

Final in-domain performance is evaluated on a block-count-balanced test set of 2,000 architectures, with 500 examples at each block count from two to five. Block-count extrapolation uses an independently generated set of 2,500 architectures, with 500 examples at each block count from six to ten. Final comparisons use five training seeds.

\subsection{RPA Stability Condition}\label{rpa-stability-condition}

The disordered-state instability of homogeneous, incompressible AB block-copolymer melts is evaluated using the established random-phase approximation (RPA) framework of Leibler and its field-theoretic formulation by Fredrickson \cite{leiblerTheoryMicrophaseSeparation1980,fredricksonEquilibriumTheoryInhomogeneous2006}. Each architecture is treated as a Gaussian chain of \(N\) statistical segments of length \(b\), with \(R_g^2=Nb^2/6\). All lengths are expressed in units of \(R_g\), such that \(R_g=1\). Accordingly, \(k\) is the dimensionless wavenumber and \(x=k^2\). Roman indices \(i\) and \(j\) label individual blocks, whereas Greek indices \(\alpha\) and \(\beta\) label chemical species in \(\{A,B\}\). Block \(i\) has volume fraction \(f_i=N_i/N\). Incompressibility leaves one independent composition-fluctuation mode \(\delta\phi(\mathbf k)\), with \(\delta\phi_A(\mathbf k)=-\delta\phi_B(\mathbf k)\).

\Needspace{12\baselineskip}
For a Gaussian chain, two segments separated by a normalized contour interval \(\Delta s\) contribute the propagator \(\exp(-x\Delta s)\). Integrating this propagator over blocks \(i\) and \(j\) gives the block-pair correlation

\[
s_{ij}(x)=
\begin{cases}
g_D(x,f_i), & i=j,\\
h(x,f_i)e^{-xd_{ij}}h(x,f_j), & i\neq j,
\end{cases}
\]

where \(\mathcal B_\alpha\) is the set of blocks of chemical species \(\alpha\), \(g_D(x,f)=\frac{2}{x^2}\left(e^{-xf}+xf-1\right)\) is the Debye function, and \(h(x,f)=\frac{1-e^{-xf}}{x}\). Because the architectures considered here are non-cyclic, any two blocks are connected by a unique path. \(d_{ij}\) is the summed contour fraction of the intervening blocks on the path connecting blocks \(i\) and \(j\), and is zero when the two blocks share a junction. Summing \(s_{ij}(x)\) over block pairs belonging to species \(\alpha\) and \(\beta\) gives the intramolecular correlation functions \(S_{\alpha\beta}(x)=\sum_{i\in\mathcal B_\alpha}\sum_{j\in\mathcal B_\beta}s_{ij}(x)\), through which the block identities, lengths, and connectivity enter the RPA calculation.

Within this formulation, stability of the disordered state is determined by expanding the dimensionless free energy in the composition fluctuation. The linear term vanishes at the homogeneous reference state, leaving

\[
\Delta F=\frac{1}{2}\sum_{\mathbf k}
\Gamma^{(2)}(x;\chi N)|\delta\phi(\mathbf k)|^2
+O(\delta\phi^3).
\]

The quadratic coefficient \(\Gamma^{(2)}\) is reciprocal to the collective composition structure factor

\[
\Gamma^{(2)}(x;\chi N)=S(x)^{-1},
\]

which can be constructed from the intramolecular correlations \(S_{\alpha\beta}(x)\) \cite{leiblerTheoryMicrophaseSeparation1980}

\[
S(x)=\frac{W(x)}{S^{\mathrm t}(x)-2\chi N W(x)},
\]

where \(S^{\mathrm t}(x)=S_{AA}(x)+S_{BB}(x)+2S_{AB}(x)\) and \(W(x)=S_{AA}(x)S_{BB}(x)-[S_{AB}(x)]^2\). The corresponding quadratic coefficient is then

\[
\Gamma^{(2)}(x;\chi N)=\frac{S^{\mathrm t}(x)}{W(x)}-2\chi N.
\]

\Needspace{10\baselineskip}
The disordered state is locally stable when \(\Gamma^{(2)}(x;\chi N)>0\) for every \(x\). As \(\chi N\) increases, the spinodal is reached when the minimum of \(\Gamma^{(2)}\) first becomes zero. Because \(-2\chi N\) is independent of \(x\), the correlation term is first minimized with respect to \(x\), after which the zero-curvature condition is imposed

\[
x^*=\underset{x>0}{\operatorname{argmin}}\,\frac{S^{\mathrm t}(x)}{W(x)}.
\]

\[
(\chi N)^*=\frac{S^{\mathrm t}(x^*)}{2W(x^*)}.
\]

\[
k^*=\sqrt{x^*}.
\]

At \((\chi N)^*\), the quadratic free-energy cost of the mode \(k^*\) vanishes and the homogeneous state becomes unstable to composition modulation. The architecture-dependent \((\chi N)^*\) and dimensionless wavenumber \(k^*\) are used as supervised targets.

\subsection{Machine learning details}\label{machine-learning-details}

\subsubsection{GNN}\label{gnn}

\textbf{Input representation.}

A shared GNN encoder processes two- to five-block architectures by updating node states through block-dependent message passing, pooling these states into a graph representation and predicting the RPA instability threshold. The model is implemented using PyTorch 2.7.1 (CUDA 11.8 build) and PyTorch Geometric 2.7.0 \cite{paszkePyTorchImperativeStyle2019,feyFastGraphRepresentation2019}.

Each polymer is represented as an undirected block--junction graph, with junctions and free ends as nodes and blocks as edges (Fig. 9a). Each physical edge supports message passing in both directions with identical attributes \([f,t]\), where \(f\) is the block volume fraction and \(t\) is the species label (1 for A and 0 for B). Each graph uses 32 node slots: \(n+1\) represents the vertices of an \(n\)-block tree, and the remaining slots are padding positions without physical edges. Nodes are initialized with trainable 512-dimensional positional vectors. Physical-node indices follow the recorded tree-growth order. Prediction sensitivity to alternative legal growth histories is assessed in Supplementary Discussion Section 3.

\textbf{Model architecture.}

The RPA response depends on intra-block, shared-junction and separated-block correlations. Along physical edges alone, information travels one hop per layer, so communication between distant nodes requires successive updates (Fig. 9c, top). To shorten these communication paths, we introduce a virtual node following the master-node construction of Gilmer \emph{et al}. \cite{gilmerNeuralMessagePassing2017}. It connects bidirectionally to all 32 slots through 64 directed edges with attributes \([0,0]\), providing a two-hop route between any pair of physical nodes (Fig. 9c, bottom). The virtual node also has a trainable 512-dimensional initial state. All 33 positions, including the virtual node and the padding slots, participate in message passing. Matched representation ablations compare physical-edge-only, self-loop, line-graph and bipartite constructions and support the selected junction--block graph with global communication (Supplementary Tables S7a--S7c and S8a--S8b).

The GNN encoder comprises five layers based on the graph isomorphism network with edge features (GINE; Hu \emph{et al}. \cite{huStrategiesPretrainingGraph2020}), each with a hidden dimension of 512. Each GINE layer combines edge-dependent neighbour messages with the central-node state (Fig. 9b).

The message and node updates are:

\[
\begin{aligned}
\mathbf m_{j\to i}^{(\ell)} &= \operatorname{ReLU}\!\left(\mathbf h_j^{(\ell)}+\mathbf E_\ell\mathbf e_{ji}+\mathbf b_\ell\right),\\
\mathbf h_i^{(\ell+1)} &= \operatorname{MLP}_\ell\!\left((1+\epsilon_\ell)\mathbf h_i^{(\ell)}+\sum_{j\in\mathcal N(i)}\mathbf m_{j\to i}^{(\ell)}\right).
\end{aligned}
\]

Here, \(\mathbf m_{j\to i}^{(\ell)}\) is the message from node \(j\) to node \(i\) at layer \(\ell\), \(\mathbf h_i^{(\ell)}\) is the node state, and \(\mathbf e_{ji}\) contains the edge attributes. The layer-specific projection parameters \(\mathbf E_\ell\) and \(\mathbf b_\ell\) map edge attributes into the hidden space, \(\epsilon_\ell\) weights the central-node contribution, and \(\mathcal N(i)\) is the set of neighbours of node \(i\) through physical or virtual edges.

After five updates, the node states are averaged to obtain the graph-level architecture representation \(\mathbf h_G\), and a single linear regression head predicts the logarithmic instability threshold:

\[
\mathbf h_G=\frac{1}{N_{\mathrm{pool}}}\sum_{i=1}^{N_{\mathrm{pool}}}\mathbf h_i^{(L)},
\qquad
\hat y=\mathbf w^{\mathsf T}\mathbf h_G+b.
\]

Here, \(L\) is the number of message-passing layers and \(N_{\mathrm{pool}}\) is the total number of pooled node positions, including physical nodes, padding positions and the virtual node. \(\mathbf w\) and \(b\) are the trainable readout parameters, and predictions are returned to the original scale as \(\widehat{(\chi N)^*}=10^{\hat y}\). The complete model contains 2,685,953 trainable parameters.

\begin{figure}[p]
\centering
\includegraphics[width=\linewidth]{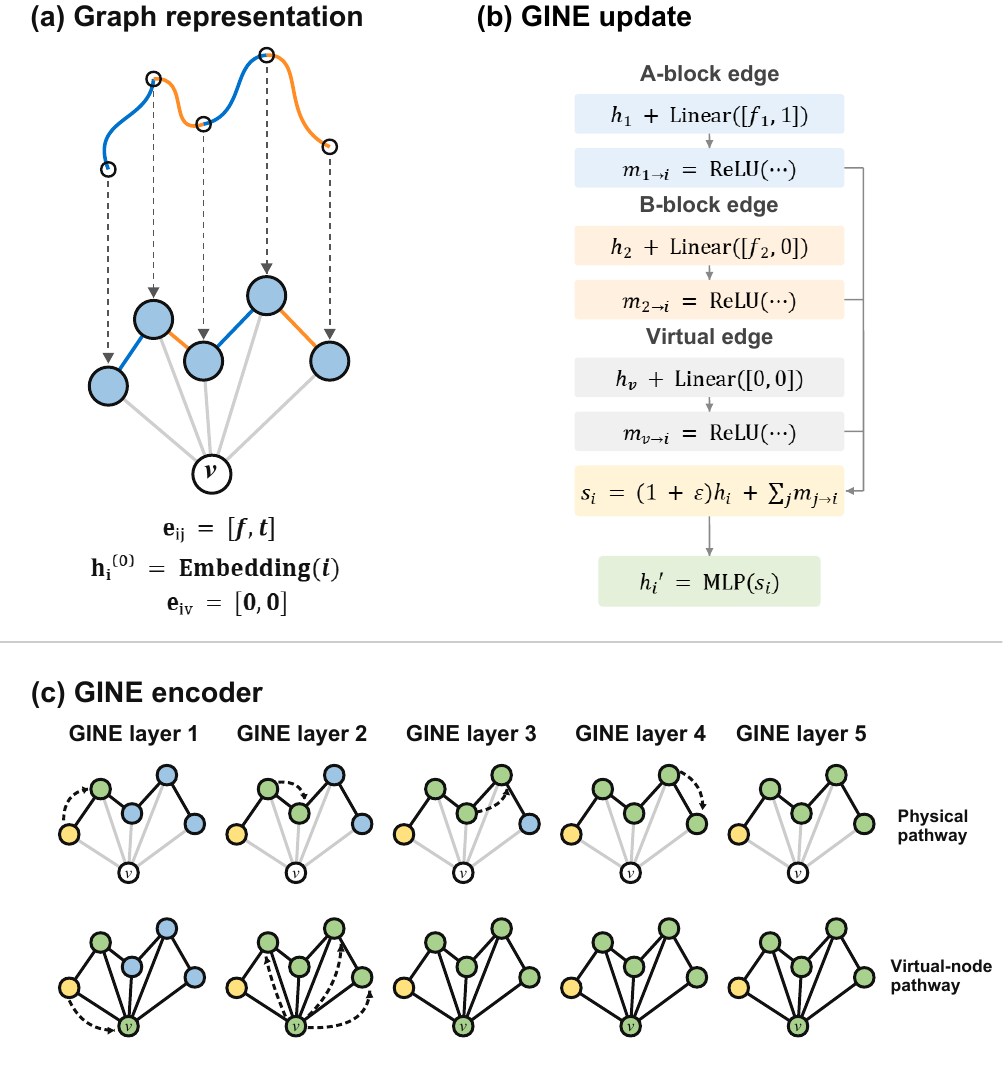}
\caption{\textbf{Graph representation and graph isomorphism network with edge features (GINE) encoder.} \textbf{a,} Mapping a polymer chain to a block--junction graph. Blue and orange edges denote A and B blocks, respectively; physical edges carry attributes \([f,t]\), node states are initialized by learnable positional embeddings, and the virtual node \(v\) is connected by edges with attributes \([0,0]\). Padding slots are omitted for clarity. \textbf{b,} A GINE update combines messages from A-block, B-block and virtual edges with the weighted central-node state, then applies an MLP. Layer indices are omitted for clarity: \(h_i\) and \(h_i'\) denote the states before and after the update, respectively, and \(\operatorname{Linear}(\mathbf e)=\mathbf E_\ell\mathbf e+\mathbf b_\ell\). \textbf{c,} Schematic comparison of propagation mechanisms across five GINE layers: information travels only along physical edges in the top row, whereas the bottom row illustrates communication assisted by the virtual node, which provides a two-hop route between any pair of physical nodes. Grey virtual connections in the top row indicate the full connectivity but are omitted from the illustrated propagation. Yellow marks the source node, green marks nodes that its information can reach along the illustrated routes, and dashed arrows indicate propagation routes. Colours mark structural reachability.}\label{fig:fig9}
\end{figure}

\textbf{Representation comparisons and hyperparameter selection.}

Models are trained with Adam at batch size 128 and initial learning rate 0.002 for at most 200 epochs, minimizing mean-squared error between predictions and targets in log space. All GNN runs use a minimum of 120 training epochs. From epoch 120 onward, early stopping is triggered after 30 epochs without a new minimum validation loss. ReduceLROnPlateau multiplies the learning rate by 0.5 after 10 epochs without improvement in validation loss.

Hyperparameters and graph representations are selected by staged one-factor-at-a-time screening and local refinement using validation MAPE across repeated seeds. After selection, the final configuration is frozen for representation and training-size comparisons. Search ranges, selection criteria and complete results are reported in Supplementary Discussion Section 6, Supplementary Tables S6--S8 and Supplementary Figs. S12--S14.

\clearpage

\subsubsection{Descriptor MLP}\label{descriptor-mlp}

\textbf{Input representation.}

The descriptor MLP baseline tests whether fixed chain-architecture features can predict the RPA response. We construct 35 candidate descriptors from the block--junction graph, combining graph statistics with composition, interface and fraction-weighted path features. These features are calculated directly from chain architecture, whose definitions are provided in Supplementary Table S9.

Handcrafted features can contain repeated information, so we screen the training fold described above for near-constant features, pairwise correlation and multivariable linear redundancy. Descriptors with variance at most \(10^{-12}\) are removed first. Candidates are then ranked by absolute Spearman correlation with the target \cite{spearmanProofMeasurementAssociation1904}. When pairwise correlation reaches \(|\rho|\geq0.95\), the more target-associated descriptor is retained, with alphabetical tie-breaking. The descriptor with the largest variance inflation factor (VIF) is subsequently removed iteratively until all retained descriptors have VIF values at most 10, adapting the descriptor-screening procedure used by Wiriyarattanakul et al. \cite{wiriyarattanakulComparativeStudyMachine2024}. The cutoff is treated here as an empirical diagnostic for linear collinearity rather than a universal exclusion rule, consistent with O'Brien's caution that VIF thresholds require contextual interpretation \cite{obrienCautionRegardingRules2007}. Calculation conventions and screening results are provided in Supplementary Discussion Section 7 and Supplementary Figs. S15--S17.

\textbf{Model architecture.}

The final MLP accepts 11 standardized descriptors and contains three fully connected hidden layers of width 256, each followed by batch normalization and ReLU, with no dropout and a linear output predicting \(\log_{10}[(\chi N)^*]\). It contains 136,449 trainable parameters. Scaling parameters are estimated from each run's training samples and applied unchanged to evaluation samples.

\textbf{Training and hyperparameter selection.}

Training uses AdamW with learning rate \(5\times10^{-4}\), weight decay \(10^{-5}\) and batch size 256, minimizing MSE of the log-transformed target. ReduceLROnPlateau halves the learning rate after eight epochs without validation-loss improvement. Training runs for at most 200 epochs with early-stopping patience 30. MAPE, RMSE, MAE and \(R^2\) are calculated after transforming predictions back to the original scale.

The final configuration is selected by five-seed validation MAPE after one-factor-at-a-time screening of batch size, learning rate, hidden-layer width, depth and dropout. Complete search ranges and comparisons are reported in Supplementary Discussion Section 7.

\textbf{Descriptor associations and model comparisons.}

Post-selection association analyses, descriptor-group ablations, individual-feature removals and the unreduced 35-descriptor control use the fixed development split and model configuration. Their methods and results are reported in Supplementary Discussion Section 7 and Supplementary Figs. S18--S21.

\subsubsection{Transformer}\label{transformer}

\textbf{Input representation.}

The canonical transformer provides a sequence-based control complementary to the descriptor MLP and the GNN model. Each architecture record represented using Polymer.jl is serialized from the block--junction graph defined above, retaining block species and fractions. Each block contributes tokens for its canonical block identifier, species, fraction, the free- or branch-end type of each endpoint, and canonical endpoint identifiers shared by blocks meeting at the same junction. A learned two-channel projection supplies the continuous fraction and a value-presence indicator. Exact graph canonicalization removes the original block and endpoint names. Sequences begin with a classification token, are padded within batches and are limited to 128 tokens. Vocabulary and maximum sequence length are fitted on the relevant fitting pool and reused without extension for internal-validation and external-validation inference. The canonicalization procedure and complete token pattern are given in Supplementary Discussion Section 8.

\textbf{Model architecture.}

The sequence encoder follows the transformer framework of Vaswani \emph{et al}. \cite{vaswaniAttentionAllYou2017}, with its application to polymer sequences motivated by TransPolymer \cite{xuTransPolymerTransformerbasedLanguage2023}. The selected transformer uses a 256-dimensional embedding, eight attention heads, three encoder layers, a 1,024-dimensional feed-forward layer, GELU, learned positional encoding and classification-token readout. It contains 2,449,921 trainable parameters.

\textbf{Training and hyperparameter selection.}

Training uses Adam for 120 epochs with learning rate \(2\times10^{-4}\), batch size 40, no dropout and weight decay 0, minimizing MSE of \(\log_{10}[(\chi N)^*]\). ReduceLROnPlateau halves the learning rate after 10 epochs without validation-loss improvement.

The final transformer configuration is selected by five-seed validation MAPE after one-factor-at-a-time screening, targeted interaction checks and local refinement. It is then frozen for the training-size analysis across the full set of nested training subsets. Complete searches and diagnostics are reported in Supplementary Tables S11 and S12 and Supplementary Figs. S22 and S23.

\textbf{Representation checks and training-size analysis.}

Canonicalization is tested independently of target values on all development-pool and external-validation records after endpoint and block renaming, endpoint reversal and input-block shuffling. With the configuration fixed, sample-size dependence is assessed under the protocol defined above; results are reported in Supplementary Discussion Section 8.

\subsection{Block-count extrapolation evaluation}\label{block-count-extrapolation-evaluation}

The validation-selected GNN, MLP and transformer models (seeds 202601--202605 for each family) are evaluated on the same OOD set using fixed parameters and their respective training-time preprocessing. GNN inference uses batches of 32, and outputs are transformed from \(\log_{10}[(\chi N)^*]\) to the original scale. For all three model families, MAPE, RMSE, MAE and \(R^2\) are calculated against the RPA reference values on the original scale and summarized as the mean and sample standard deviation across the five models. Figure 4 reports the block-count-resolved five-model means and sample standard deviations.

\subsection{Mechanistic interpretability analyses}\label{mechanistic-interpretability-analyses}

All three analyses use the frozen five-layer GNN model and the topology \(A_1B_1(B_2,B_3)_TA_2\), with \(f_{A_1}=f_{A_2}=f_A/2\), \(f_{B_1}=(1-f_A)\rho/(\rho+2)\) and \(f_{B_2}=f_{B_3}=(1-f_A)/(\rho+2)\). Combining nine values \(f_A=0.1,0.2,\ldots,0.9\) with five ratios \(\rho=0.5,1,2,5,10\) defines 45 block-fraction parameterizations of this fixed topology. All RPA correlation components are evaluated at the instability wavenumber \(k^*\) of the corresponding sample.

\textbf{RPA edge influence and message deletion.}

At each sample's numerically determined \(k^*\), block-pair contributions \(s_p\) to \(S_{AA}\), \(S_{BB}\) and \(S_{AB}\) are classified as self, shared-junction or separated. The sensitivity of the spinodal objective to channel \(\alpha\beta\) is defined as \(g_{\alpha\beta}=\partial[S^{\mathrm t}/(2W)]/\partial S_{\alpha\beta}\). Each sensitivity-weighted pair contribution is distributed uniformly over its unique block-edge path \(P_p\):

\[
I_{e,c}^{\rm RPA}=
\sum_{\alpha\beta\in\{AA,BB,AB\}}
\sum_{\substack{p\in\mathcal P_{\alpha\beta,c}\\ e\in P_p}}
g_{\alpha\beta}(x^*)\,
\frac{\mu_p s_p(x^*)}{|P_p|}.
\]

Summing over the three relation classes gives the total influence on edge \(e\):

\[
I_e^{\rm RPA}=\sum_{c\in\{\rm self,junction,separated\}}I_{e,c}^{\rm RPA}.
\]

Here, \(\mathcal P_{\alpha\beta,c}\) contains the block pairs in channel \(\alpha\beta\) and relation class \(c\), and \(|P_p|\) is the number of path edges. The ordered-pair multiplicity is \(\mu_p=2\) for distinct AA or BB pairs and \(\mu_p=1\) for self and AB pairs. Thus, \(I_e^{\rm RPA}\) is a signed edge-level measure obtained by allocating sensitivity-weighted correlation contributions at fixed \(x^*\).

Model dependence on edge \(e\) is measured by suppressing both reciprocal directed messages associated with that physical edge while retaining all other physical and virtual-edge messages:

\[
F_e=\left|\hat y(\mathbf 1)-\hat y(\mathbf m^{(-e)})\right|,
\]

Here, \(\mathbf m^{(-e)}\) suppresses the two directed messages, and \(\hat y\) is the frozen prediction on the original \((\chi N)^*\) scale. For each structure, \(F_e\) is averaged across three legal node relabellings. We compare \(|I_e^{\rm RPA}|\) with mean \(F_e\) using Spearman correlation across 225 edge pairs and within each architecture instance (Supplementary Discussion Section 3.1).

The channel sensitivities \(g_{\alpha\beta}\) are evaluated analytically at each sample's \(x^*\). Their explicit expressions are provided in Supplementary Discussion Section 3.1.

\textbf{Layerwise RPA concept probes.}

Separate PCA--ridge probes relate the 512-dimensional global-mean-pooled representations at the input (\(l=0\)) and each GINE layer (\(l=1,\ldots,5\)) to \(y_i\in\{S_{AA}(k^*),S_{BB}(k^*),S_{AB}(k^*),k^*,(\chi N)^*\}\). Each probe comprises standardization, projection onto two whitened principal components and ridge regression with penalty 10. Five-fold cross-validation holds out one complete \(\rho\) group per fold, giving 36 training and 9 evaluation samples. All preprocessing and regression are fitted within the training fold, with fixed hyperparameters and no GNN updates. Probe performance is measured by \(R^2\) from the pooled out-of-fold predictions for all 45 samples.

\textbf{Block-placement counterfactuals.}

Within each \((f_A,\rho)\) group, the five \((\mathrm{species},f)\) attributes are permuted over fixed block-edge positions, preserving topology, degree sequence, total composition and the attribute multiset while changing shared-junction relations and intervening contour distances. Layerwise graph embeddings are averaged over legal growth histories, and embedding and RPA-vector distances are compared over 16,065 matched pairs. Grid-wise correlations, assignment counts and ordering-sensitivity analyses are reported in Supplementary Discussion Section 3.

The matched-pair statistics are defined as follows:

\[
\overline{\mathbf h}_i^{(l)}=\frac{1}{|\mathcal H_i|}\sum_{\pi\in\mathcal H_i}\mathbf h^{(l)}(G_i^{\pi}).
\]

The RPA vector for each assignment is:

\[
\mathbf c_i=\left[S_{AA,i}^{\rm separated}(k_i^*),S_{BB,i}^{\rm separated}(k_i^*),S_{AB,i}^{\rm separated}(k_i^*),k_i^*\right].
\]

The embedding distance at layer \(l\) is:

\[
d_{ij}^{(l)}=\left\|\overline{\mathbf h}_i^{(l)}-\overline{\mathbf h}_j^{(l)}\right\|_2.
\]

The corresponding RPA distance is:

\[
\delta_{ij}^{\rm RPA}=\left\|\mathbf c_i-\mathbf c_j\right\|_2.
\]

Their rank association over matched pairs is:

\[
\rho_s^{(l)}=\operatorname{Spearman}\!\left(\{d_{ij}^{(l)}\}_{(i,j)\in\mathcal M},\{\delta_{ij}^{\rm RPA}\}_{(i,j)\in\mathcal M}\right).
\]

Here, \(G_i^{\pi}\) is assignment \(i\) serialized under legal growth history \(\pi\), \(\mathcal H_i\) is its set of legal histories and \(\mathcal M\) contains distinct assignment pairs within the same \((f_A,\rho)\) cell.

\subsection{Cross-target transfer learning protocols}\label{cross-target-transfer-learning-protocols}

The five protocols differ in encoder initialization, the encoder-update schedule and prediction-head capacity.

We compared five protocols for \(k^*\) prediction: scratch training, frozen-encoder linear decoding, frozen-embedding MLP decoding, end-to-end fine-tuning and two-stage transfer. Scratch training randomly initialized the encoder and a linear prediction head. Frozen-encoder linear decoding kept a \((\chi N)^*\)-pretrained encoder fixed and trained a new linear prediction head, whereas frozen-embedding MLP decoding trained a nonlinear multilayer perceptron (MLP) on fixed vectors from the same encoder. End-to-end fine-tuning updated the pretrained encoder and a new linear prediction head from epoch 1. Two-stage transfer updated only the linear prediction head during epochs 1--50, then jointly updated the encoder and head during epochs 51--200.

For each of the five random seeds (202601--202605), each transfer run loaded the \((\chi N)^*\) encoder pretrained with the same seed. Scratch training, frozen-encoder linear decoding, end-to-end fine-tuning and two-stage transfer used the learnable-node, virtual-node graph architecture, a batch size of 32 and encoder and linear-head learning rates of \(5\times10^{-4}\) and \(5\times10^{-3}\). Adam used ReduceLROnPlateau with a factor of 0.5 and a patience of 10 epochs. Training ran for 200 epochs without early stopping and retained the checkpoint with the lowest validation MSE of \(\log_{10}(k^*)\).

Frozen-embedding MLP decoding fitted an independent 512--256--128--1 MLP to fixed 512-dimensional global-mean-pooled vectors from each pretrained encoder. Inputs were standardized using the training-subset statistics, and each hidden layer used LayerNorm, ReLU and dropout of 0.1. AdamW used a learning rate of \(10^{-3}\), weight decay of \(10^{-4}\) and a batch size of up to 512. Training was limited to 500 epochs, with early stopping after 40 epochs without improvement in validation loss.

Limited-data transfer used strictly nested, block-count-balanced subsets of 100, 250, 500, 1,000 and 2,000 \(k^*\)-labelled architectures. Within each seed, all protocols shared the same training subsets and a paired 1,000-architecture validation subset containing 250 examples for each block count from two to five. The source encoders had encountered these architectures only with \((\chi N)^*\) labels. Full-data transfer used the complete training set defined above. Predictions were inverse-transformed before metric calculation, and all other procedures followed the common protocol described above.

\backmatter

\bookmarksetup{startatroot}
\bmhead{Data availability}

The datasets used in this study are available on Zenodo at \url{https://doi.org/10.5281/zenodo.22685926}.

\bookmarksetup{startatroot}
\bmhead{Code availability}

The source code used to generate the architectures and RPA labels, train and evaluate the models, and reproduce the reported analyses is publicly available at \url{https://github.com/PolyorderLab/BCP-GraphRPA}.

\bibliography{bib}

\bookmarksetup{startatroot}
\bmhead{Acknowledgements}

This work was supported by the National Natural Science Foundation of China (Grant No. 21873021).

\bookmarksetup{startatroot}
\bmhead{Author contributions}

X.Z. led the research and performed the majority of the methodology development, software implementation, data analysis, model evaluation, and visualization, and wrote the original manuscript. Y.W. designed and implemented the descriptor-MLP component of the study. Y.L. conceived and supervised the study, contributed to the interpretation of the results, and reviewed and edited the manuscript. All authors approved the final manuscript.

\bookmarksetup{startatroot}
\bmhead{Competing interests}

The authors declare no competing interests.

\bookmarksetup{startatroot}
\bmhead{Supplementary information}

Supplementary information accompanies this paper and includes additional in-domain and block-count extrapolation results, mechanistic analyses, transfer-learning evaluations, model-selection results, representation ablations and descriptor definitions.

\end{document}